# Phase Noise Influence in Coherent Optical OFDM Systems with RF Pilot Tone: Digital IFFT Multiplexing and FFT Demodulation

Gunnar Jacobsen[1,*], Tianhua Xu[1,2,3], Sergei Popov[2], Jie Li[1], Ari T. Friberg[2] and Yimo Zhang[3]

[1]Acreo AB, Electrum 236, SE-16440, Kista, Sweden

[2]Royal Institute of Technology, Stockholm, SE-16440, Sweden

[3]Tianjin University, Tianjin, 300072, P.R. China

**Abstract**: We present a comparative study of the influence of dispersion induced phase noise for CO-OFDM systems using Tx channel multiplexing and Rx matched filter (analogue hardware based); and FFT multiplexing/IFFT demultiplexing techniques (software based). An RF carrier pilot tone is used to mitigate the phase noise influence. From the analysis, it appears that the phase noise influence for the two OFDM implementations is very similar. The software based system provides a method for a rigorous evaluation of the phase noise variance caused by Common Phase Error (CPE) and Inter-Carrier Interference (ICI) and this, in turns, leads to a BER specification. Numerical results focus on a CO-OFDM system with 1GS/s QPSK channel modulation. Worst case BER results are evaluated and compared to the BER of a QPSK system with the same capacity as the OFDM implementation. Results are evaluated as a function of transmission distance, and for the QPSK system the influence of equalization enhanced phase noise (EEPN) is included. For both types of systems, the phase noise variance increases significantly with increasing transmission distance. An important and novel observation is that the two types of systems have very closely the same BER as a function of transmission distance for the same capacity. For the high capacity QPSK implementation, the increase in BER is due to EEPN, whereas for the OFDM approach it is due to the dispersion caused walk-off of the RF pilot tone relative to the OFDM signal channels. For a total capacity of 400 Gb/s, the transmission distance to have the BER < $10^{-4}$ is less than 277 km.

For an RF pilot located in the center of the OFDM band in a CO-OFDM implementation with n-level PSK channel modulation the current results suggest that the walk-off effect is equivalent to the EEPN impact in a single channel n-level PSK system with the same capacity. This observation is important for future design of coherent long-range systems since it shows that there is a free choice between CO-OFDM and a high capacity nPSK implementation at least as long as the phase noise influence is concerned.

## 1. Introduction

Coherent optical communications research today has focus on achieving high capacity system bit-rates (100 Gb/s – 1 Tb/s) with the possibility of efficient optical multiplexing (MUX) and demultiplexing (DEMUX) on sub-band level (the order of 1 Gb/s). An essential part of the optical system design is the use of Discrete Signal Processing (DSP) techniques in both transmitter and receiver to eliminate costly hardware for dispersion compensation, polarization tracking and control, clock extraction, carrier phase extraction etc.

In the core part of the network, emphasis has been on long-range (high sensitivity) systems where coherent (homodyne) implementations of n-level Phase-Shift-Keying (nPSK) and Quadrature Amplitude Modulation (nQAM) have proven superior performance. When it comes to efficient high-capacity and low granularity, optical MUX/DEMUX Orthogonal Frequency Division Multiplexing (OFDM) technology becomes an interesting alternative. The MUX/DEMUX capability is of special interest in the Metro-/Access parts of the optical network where long transmission range is not a prime factor. OFDM systems can be viewed as a sub-carrier multiplexed optical system and, due to the need of a strong "DC" optical carrier wave (in order to avoid clipping distortion effects), these systems will have lower sensitivity requirements (shorter reach) than nPSK or nQAM systems with equivalent capacity [1]. However, OFDM systems have other advantages due to the distributed capacity in many tightly spaced signal channels in the frequency domain. These advantages include highly efficient optical reconfigurable optical networks (efficient optical MUX/DEMUX), easy upgrade of transmission capacity using discrete (digital) software (Digital Inverse Fast-Fourier-Transform (DIFFT) can be used for channel MUX and DFFT for channel DEMUX) and adaptive data provisioning in the optical domain on a per OFDM-channel basis (i.e. optical ADSL implementation to make transmission agnostic to underlying physical link).

* **Corresdonding author:** Gunnar Jacobsen, Acreo AB, Electrum 236, SE-16440 Kista, Sweden. E-mail: gunnar.jacobsen@acreo.se.

Optical coherent systems can be seen as a complementary technology to modern systems in the radio (mobile) domain. It is important to understand the differences in these implementations and these are mainly that the optical systems operate at significantly higher transmission speeds than their radio counterparts and that they use signal sources (transmitter and local oscillator lasers) which are significantly less coherent than radio sources. For nPSK and nQAM systems, DSP technology in the optical domain is entirely focused on high speed implementation of simple functions, such as AD/DA currently operating at 56 Gbaud [2]. The use of high constellation transmission schemes is a way of lowering the DSP speed relative to the total capacity. Using OFDM as MUX/DEMUX technology and implementing hundreds or thousands channels is an alternative approach of very efficient lowering the DSP speed (per channel) and still maintaining 100 Gb/s (or more) total system throughput. Both Direct Detection and Coherent (heterodyne) detection is considered for OFDM implementations (DD-OFDM and CO-OFDM systems) and the relatively low channel baud-rate leads to an influence of phase noise which can be severe.

The theory basis for dealing with the phase noise influence has been presented for radio OFDM systems in [3-8], and the option accounting for optical systems can be found in [9-15]. DD-OFDM optical systems are considered specifically in [12-15], CO-OFDM optical systems are considered in [9-11, 15]. The special DD-OFDM radio-over-fiber system operating in the 60 GHz radio band is analyzed in [14]. Nonlinear amplification and phase noise for radio OFDM systems are considered in [16].

Using nPSK or nQAM systems with DSP based dispersion compensation leads to strong influence of laser phase noise which is further enhanced by equalization enhanced phase noise (EEPN) originating from the local oscillator laser [17-19]. OFDM systems may use wrapping of the signal in the time domain (cyclic prefix) to account for dispersion effects in this way eliminating the need for DSP based compensation. Using an RF pilot carrier which is adjacent to or part of the OFDM channel grid is an effective method of eliminating the phase noise effect [9, 12-15], but it has to be noted that the dispersion influenced delay of OFDM channels will make the elimination incomplete. This leads to a transmission length dependent (dispersion enhanced) phase noise effect [12-15]. In contrast, it is worth to mention that for nPSK and nQAM implementations the RF pilot carrier may eliminate the phase noise entirely. However it is important to note that the EEPN cannot be eliminated [20]. We specifically emphasize that OFDM systems do not employ electronic CD compensation and thus EEPN is not a significant effect to consider in the practical system design. It can be seen that for the same channel baud rate and total OFDM system capacity the largest phase noise walk-off appears for DD-OFDM systems and, thus, these systems are more influenced by phase noise than CO-OFDM systems [15].

System simulations (transmission experiments implemented in a software environment) have proven to be efficient design tools for nPSK/nQAM systems using partly university developed system models [17, 18] and partly commercial simulation tools [21]. Such simulations, e.g. the bit-error-rate (BER) are possible because practical system implementations are now based on forward-error-correction (FEC) where a “raw” BER (without FEC) of the order of $10^{-3}$ is sufficient. For OFDM with hundreds or thousands of signal channels, it is obvious that direct simulation of the OFDM system BER with independent simulation data (PRBS sequences) for each signal channel is a formidable task which is difficult for realization even for modern computers. Thus, it is of special interest for OFDM system models to develop insight based upon rigorous analytical models for important system parts.

It has to be pointed out that the phase noise analysis in [3-5, 15] assumes a matched filter receiver implementation whereas a FFT demux and detection method is the basis for the analysis in [6-14, 16]. The matched filter detection OFDM (based on a classical analogue hardware) and the FFT (using modern software) implementations are two interesting alternatives for the practical system which are very worthwhile to compare. The purpose of this paper is to investigate in detail the basic phase noise sensitivity for the FFT implementation on a novel analytical basis for CO-OFDM systems with RF pilot tone phase noise compensation. The developed theory will provide example results for the phase noise sensitivity. These results will be compared to the results for the CO-OFDM matched filter systems from [15] and to the results for a single channel high capacity QPSK system.

Based on previous, more approximate analysis in [15], it is seen that the CO-OFDM system will give the longest system range as well as the least phase noise sensitivity when compared to a DD-OFDM implementation. In this paper we will focus on the CO-OFDM system.

# 2. System modeling and theory

Here we display layouts for CO-OFDM systems using classical analogue subcarrier MUX and DEMUX with matched filter detection (Figure 1), and using IFFT MUX and FFT DEMUX in a software based system implementation (Figure 2).

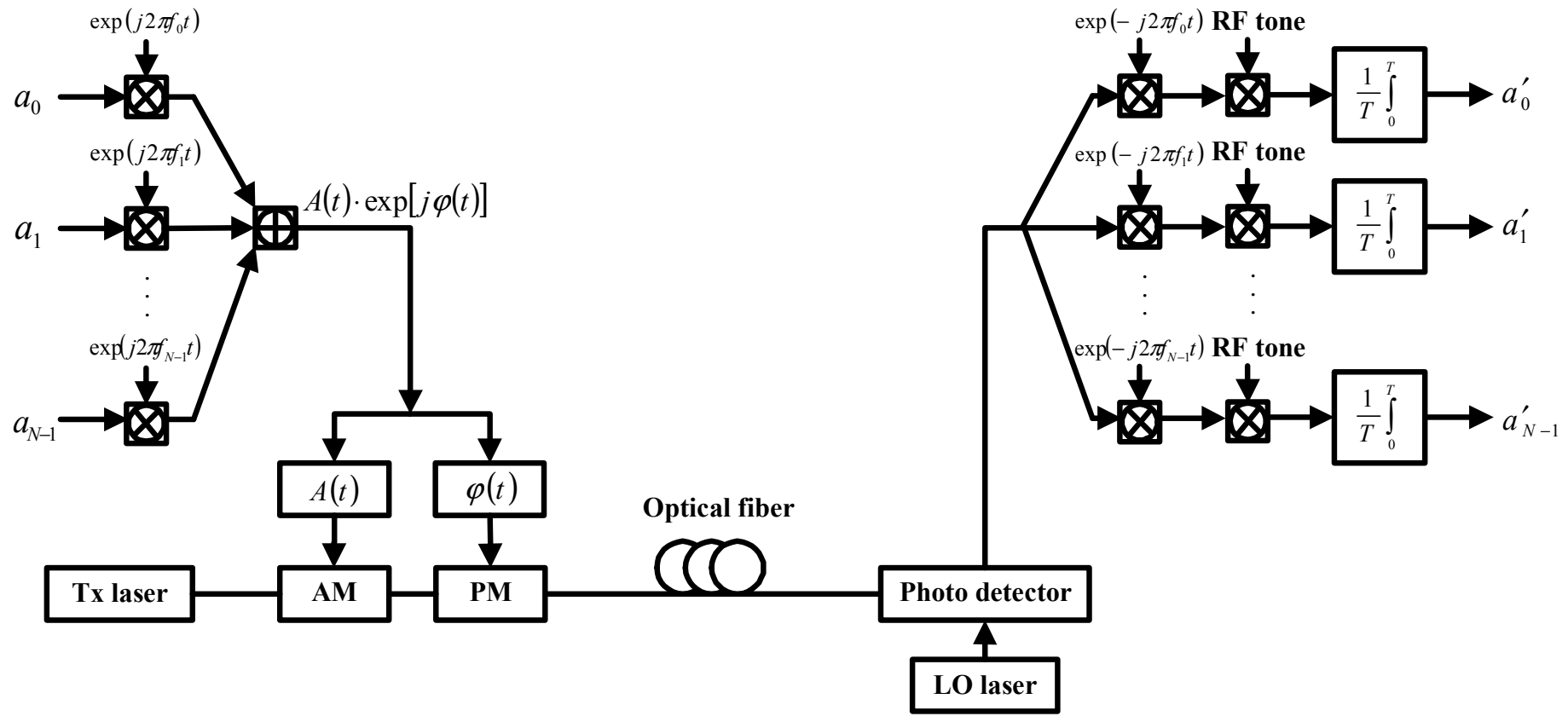

**Figure 1.** OFDM system with classical subcarrier MUX and matched filter detection including RF pilot tone phase noise mitigation. The mathematics for the MUX and DEMUX operation is schematically indicated and discussed in detail in the text. Figure abbreviations: $a_0$-$a_{N-1}$ – constellation of $N$ transmitted OFDM symbols; $a'_0$-$a'_{N-1}$ – constellation of $N$ received OFDM symbols; $f_0$-$f_{N-1}$ – OFDM channel frequencies; AM – amplitude modulator; PM – phase modulator, Tx – transmitter, LO – local oscillator; RF – radio frequency.

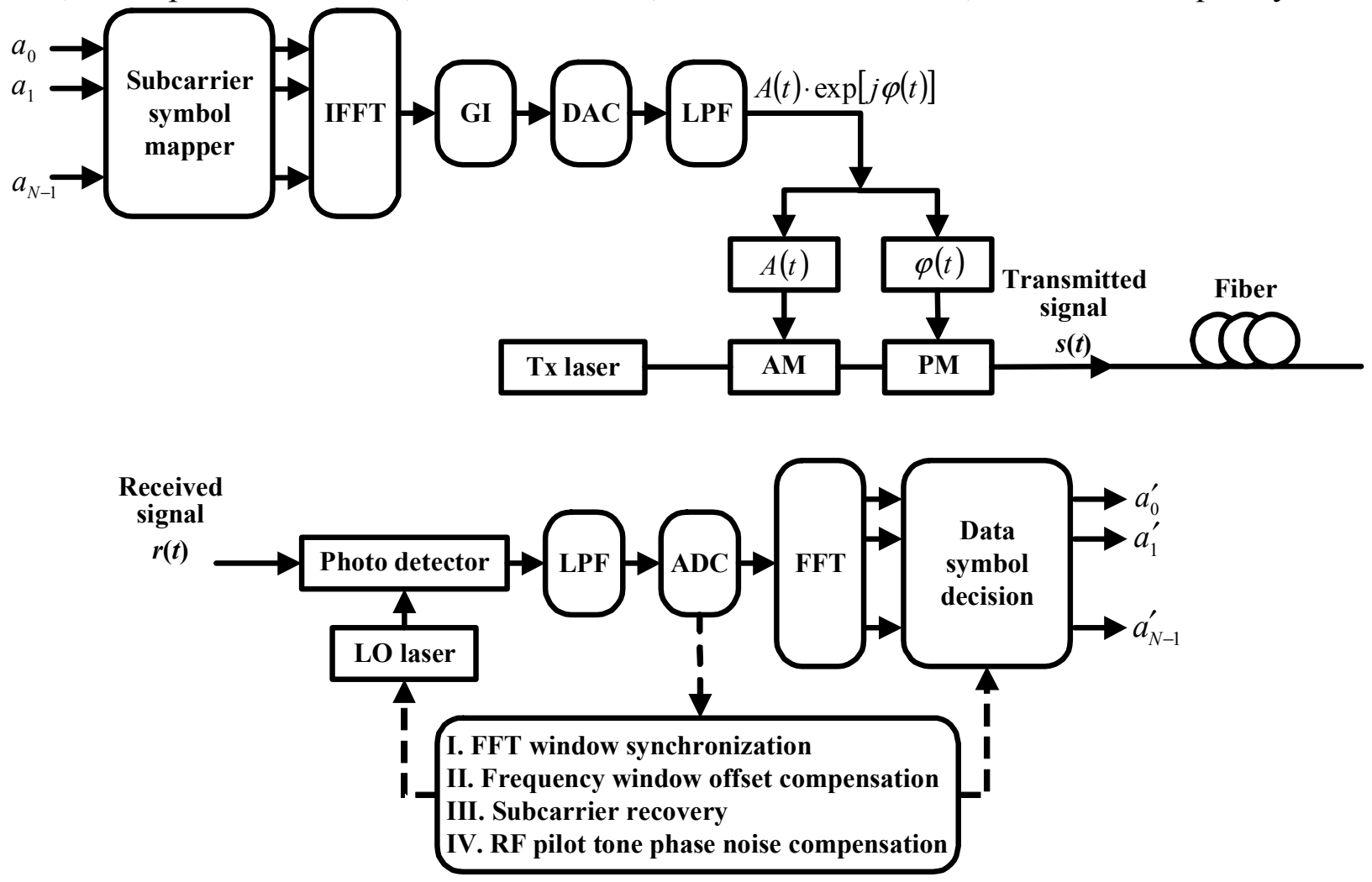

**Figure 2.** OFDM system including IFFT MUX and FFT with an RF pilot tone for phase noise mitigation. The mathematics for the MUX and DEMUX is schematically indicated and discussed in detail in the text. Figure abbreviations: $a_0$-$a_{N-1}$ – constellation of $N$ transmitted OFDM symbols; $a'_0$-$a'_{N-1}$ – constellation of $N$ received OFDM symbols; IFFT – Inverse Fast Fourier Transform; GI – guard time insertion; DAC – discrete to analogue conversion; LPF – low pass filter; AM – amplitude modulator; PM – phase modulator, Tx – transmitter, LO – local oscillator; RF – radio frequency; ADC – analogue to discrete conversion; FFT – Fast Fourier Transform.

## 2.1 CO-OFDM system with matched filter detection

In the following, we will present the derivation for CO-OFDM systems with classical matched filter detection explicitly. During a symbol period $T$ the complex envelope (constellation position) of one of the $N$ transmitted OFDM signal (defined as shown in Figure 1) is $a_k$ ($k=0,1,\ldots,N-1$). Symbol of number $k$ is moved to the electrical carrier frequency $f_k=k/T$. The $N$ symbols are multiplexed (added), and the multiplexed signal is denoted $A(t)\cdot exp(j(\varphi(t))$. The multiplexed signal is put onto the optical carrier wave and the resulting signal in the optical domain is:

$$s(t) \equiv A(t)\cdot \exp\left(j\left(2\pi f_o t + \psi(t) + \varphi(t)\right)\right) = e^{j(2\pi f_o t + \psi_{Tx}(t))} \sum_{\substack{k=0 \\ k\neq (N-1)/2}}^{N-1} a_k e^{j2\pi \frac{k}{T} t} \tag{1}$$

where $\psi_{Tx}(t)$ denotes the Tx laser phase noise and $f_o$ the optical carrier frequency. In the following, we assume, for convenience and without any loss of generality, that $N$ is odd, i.e. when the center of the OFDM grid is used for the RF carrier we have $(N-1)/2$ channels at frequencies above and below the RF carrier. We note for later use (in section 2.2) that the electrically multiplexed signal is the analogue output after digital Inverse Fast Fourier Transform (IFFT) of the digitized input sampled with $N$ bins separated by $T/N$, and each sample specifying one OFDM channel constellation $a_k$. The RF pilot carrier is injected into the analogue signal at grid position $k=(N-1)/2$ prior to optical modulation that brings $s(t)$ onto the optical carrier wave [5] (Figure 1). After coherent detection with a local oscillator (LO) laser with the same carrier frequency as the Tx laser, the output of the receiver, including correlation detection but without using the RF pilot carrier, is for symbol $k$ ($0 \le k \le N-1$) [1]:

$$a'_k = \frac{e^{-j2\pi\psi_{LO}(t)}}{T}\int_0^T s(t)\exp\left(-j2\pi\left(\frac{k}{T}+f_o\right)t\right)dt \tag{2}$$

where $\psi_{LO}(t)$ denotes the LO laser phase noise. In the case of no phase noise influence, the orthogonality between the channels means that $a'_k=a_k$ and the symbol detection is perfect. In the case of using the RF pilot carrier to minimize the phase noise influence (by complex conjugation operation as part of the data symbol identification in the Rx and using the RF pilot carrier frequency in the center of the OFDM grid for phase noise cancellation [10-15]) the influence of the LO phase noise is cancelled in (2). Taylor expansion is now employed to identify the leading order phase noise influence in (2). The resulting Common Phase Error (CPE) for channel $k$ is:

$$\frac{j}{T}\int_0^T\left(\psi_{Tx}(t)-\psi_{Tx}\left(t+\left[\frac{N-1}{2}-k\right]\cdot\tau\right)\right)dt \equiv \int_0^T \Delta\psi_{Tx,k}(t)dt \tag{3}$$

where $\tau = DL\lambda^2\Delta f / c$ ($D$ is the fiber dispersion coefficient, $L$ - the fiber length, $\lambda$ - the laser transmission wavelength, $\Delta f$ - the frequency separation between OFDM channels and $c$ is the speed of light) is specifying the dispersion influence (between adjacent OFDM channels). The Inter-Carrier Interference (ICI) is:

$$\frac{j}{T}\sum_{\substack{l\neq k\\ l=0}}^{N-1} a_l \int_0^T \Delta\psi_{Tx,l}(t)\cdot\exp\left(j\frac{2\pi(k-l)}{T}t\right)dt \tag{4}$$

The use of a common RF pilot tone in the system [5, 8], which is complex conjugated and multiplied with the OFDM signal channels, is modeled as providing a common phase reference. This eliminates the phase noise influence which is not due to dispersion for the CPE and the ICI. Since the Local Oscillator phase noise is not influenced by fiber dispersion it is completely eliminated by the RF pilot tone.

## 2.2 CO-OFDM system with IFFT MUX and FFT DEMUX and detection

A system diagram for a CO-OFDM system employing IFFT MUX and FFT DEMUX and detection operation is presented in Figure 2.

To derive the signal representation, we follow a procedure as in section 2.1. We consider an ideal system and neglect the influence of the guard time insertion, and assume that the subcarrier recovery is perfect in the following analysis. It appears directly that the signal in the optical domain is also given in this case as (1) for the discrete electrical signal sampled $N$ times during an OFDM symbol period $T$, i.e. for $t=nT/N$ ($n=0,1,\ldots,N-1$). When investigating the phase noise influence, we initiate our derivations from [6-8, 13, 14] with specific consideration of the CO-OFDM system implementation with an RF pilot tone for the phase noise mitigation and with direct influence of the fiber dispersion. This leads to an expression for the CPE that can be given as

$$\frac{j}{N}\sum_{m=0}^{N-1}\left[\psi_{Tx}\left(\frac{mT}{N}\right)-\psi_{Tx}\left(\frac{mT}{N}+\left(\frac{N-1}{2}-k\right)\tau\right)\right] \equiv \frac{j}{N}\sum_{\substack{m=0\\ m\neq(N-1)/2}}^{N-1}\left[\Delta\psi_{Tx,k}\left(\frac{mT}{N}\right)\right] \tag{5}$$

Similarly, the ICI is now given as ($a_{(N-1)/2} \equiv 1$)

$$\begin{aligned}&\frac{j}{N}\sum_{\substack{r=0\\ r\neq k}}^{N-1} a_r \sum_{m=0}^{N-1}\left[\psi_{Tx}\left(\frac{mT}{N}\right)-\psi_{Tx}\left(\frac{mT}{N}+\left(\frac{N-1}{2}-r\right)\tau\right)\right]\cdot\exp\left(\frac{j2\pi(r-k)m}{N}\right)\\ &\equiv \frac{j}{N}\sum_{\substack{r=0\\ r\neq k}}^{N-1} a_r \sum_{m=0}^{N-1}\left[\Delta\psi_{Tx,r}\left(\frac{mT}{N}\right)\right]\cdot\exp\left(\frac{j2\pi(r-k)m}{N}\right)\end{aligned} \tag{6}$$

One can note, as a novel observation, that (5) is a discrete approximation of (3), and (6) is a discrete approximation of (4), and that the approximation is becoming more accurate as $N$ (the number of OFDM symbols (and OFDM channels)) becomes large. The above derivation of (5) and (6) is in agreement with results in [6-8], but represents an extension because the influence of an RF pilot carrier is included. Note that when the CPE and ICI influence is connected to the phase detection which effects both nPSK and nQAM systems, it is associated to the imaginary parts of (5)-(6) ((5) is purely imaginary). The resulting phase noise influence on the OFDM system performance (for instance specified through the resulting phase noise variance and the associated bit-error-rate floor ($BER_{floor}$) position) can be derived from (5) and (6) on a more rigorous basis than using (3) and (4) (as it was done in [15]). This is the case because (3) and (4) include integration over the stochastic phase noise variable (which is a complicated mathematical task, see e.g. [22, 23]) whereas this is not the case for (5) and (6).

We will derive the phase noise variance associated with phase detection using (5)-(6) in two limiting cases of special interest for optical CO-OFDM systems, namely when 1) $N\tau$ and $T$ is of the same order of magnitude; and 2) $T > N\tau$. These derivations represent novel results of the combined fiber dispersion/phase noise influence for such systems. In the situation where $N\tau >> T$ the use of an RF tone for phase noise compensation is clearly not effecient.

We first note that in (6) it is sensible (in a worst case sense) to assume full correlation between the phase noise samples of the different OFDM channels which are detected at any given time instant (i.e. for a time $mT/n$ there is correlation between all phase noise samples $\Delta\psi_{Tx,l}(mT/N)$ for $l=0,1,...,N-1$). Depending on the relation between $T$ and $\tau$ the phase noise samples for different times may also be correlated. The contribution of the different phase noise samples in (5) and (6) to the total CPE+ICI phase noise variance is now taking into account 1) that each phase noise sample $\Delta\psi_{Tx,l}(mT/N),\ l=0,1,...,N-1$ is a Gaussian zero-mean stochastic variable with variance $\sigma^2_l = 2\pi\cdot\Delta\nu_{Tx}\cdot|l-(N-1)/2|\cdot\tau$ and 2) that phase noise variance contributions from two uncorrelated phase noise samples are added on phase noise variance basis whereas contributions from two fully correlated samples (numbers $l$ and $m$) are added on square-root-variance basis ($\sigma^2_{l+m} = (\sigma_l + \rho\sigma_m)^2$ with correlation $\rho = 1$ or $\rho = -1$).

In the case of $\tau >> T$ we have strong influence of fiber dispersion (corresponding to relatively long haul transmission), and the differential phase noise contributions for different times (different $m$-values) to the summations in (5) and (6) are fully correlated. This means that the CPE and ICI parts of the phase noise influence in (5) and (6) must be analyzed together, and we find the total phase noise variance:

$$\sigma^2_{k,CPE+ICI} \approx \frac{2\pi\Delta\nu_{Tx}\tau}{N^2}\left[\sum_{m=0}^{N-1}\sum_{r=0}^{N-1}\sqrt{\left|r-\frac{N-1}{2}\right|}\times \mathrm{Re}\left(\frac{a_r}{a_k}\exp\left(j\frac{2\pi(r-k)m}{N}\right)\right)\right]^2 \tag{7}$$

The contributions for different interfering channels add on a field basis and this may (depending on the symbol constellation) result in a high value for the total phase noise variance.

In the second limiting case with $T >> \tau$ we have a small influence of the fiber dispersion and because of this the differential phase noise contributions in the summations of (5) and (6) are uncorrelated for different time samples (different $m$-values in (5)-(6)). However, it is important to note that phase noise samples originating from different channel locations (different $r$-values) are fully correlated. This means that the CPE and the ICI contrtibution to the phase noise variance need to be accounted for at the same time. The following expression for the resulting phase noise variance appears:

$$\sigma^2_{k,CPE+ICI} \approx \frac{2\pi\Delta\nu_{Tx}\tau}{N^2}\sum_{m=0}^{N-1}\left[\sum_{r=0}^{N-1}\sqrt{\left|r-\frac{N-1}{2}\right|}\times \mathrm{Re}\left(\frac{a_r}{a_k}\exp\left(j\frac{2\pi(r-k)m}{N}\right)\right)\right]^2 \tag{8}$$

It is of significance for the situations specified in (7) and (8) that the ICI part of phase noise depends on the detected symbols and it is stronger when the detected symbol (in i.e. a QAM constellation) is close to the center, and interfering symbols have constellations with larger magnitude.

It is possible to derive the phase noise variance in exact form accounting in detail for the partial phase noise correlation between different channel locations in the OFDM frame (for $T>>\tau$). This can be done by introducing the correlation coefficient between two time-overlaping Wiener processes $s$ and $r$. They have the correlation coefficient $\rho_{s,r}=(min(\sigma^2_s, \sigma^2_r)/ max(\sigma^2_s, \sigma^2_r))^{1/2}$ with $\rho_{s,s}=\rho_{r,r}=1$. Then (8) is modified to read

$$\begin{aligned}\sigma^2_{k,CPE+ICI} = \frac{2\pi\Delta\nu_{Tx}\tau}{N^2}\sum_{m=0}^{N-1}\sum_{r=0}^{N-1}\sum_{s=0}^{N-1}&\left[\sqrt{\left|r-\frac{N-1}{2}\right|}\times \mathrm{Re}\left(\frac{a_r}{a_k}\exp\left(j\frac{2\pi(r-k)m}{N}\right)\right)\right.\\ &\left.\times\rho_{r,s}\sqrt{\left|s-\frac{N-1}{2}\right|}\times \mathrm{Re}\left(\frac{a_s}{a_k}\exp\left(j\frac{2\pi(s-k)m}{N}\right)\right)\right]\end{aligned} \tag{9}$$

We note that the time correlation between contributions from neighboring channels is strong ($\rho_{s,r} \approx 1$ in this case). As a sanity check it is observed that assuming full correlation ($\rho_{s,r} \equiv 1$ for all $s$ and $r$ values) makes (9) equal to (8).

We will investigate the resulting phase noise variance in more detail in the numerical examples of the next section.

When considering the amplitude of the phase noise contribution which influences detection of the length (magnitude) of $a_k$, there is no contribution from the CPE part of the phase noise as can be seen from (5). The ICI part will, in the limit of $\tau >>T$, give a contribution (from the real part of (6)) which can be specified in similar forms as (7) and (8).

We note that practical nPSK, as well as nQAM, systems can be designed by choosing constellation configurations such that the phase noise influence on the detected phase is the dominating phase noise contribution. In the following, we will not consider the magnitude part of the phase noise influence.

## 3. Simulation results and discussion

It is of interest to compare the normalized (dividing by the intrinsic phase noise variance $2\pi\Delta\nu_{Tx}\tau$) CPE+ICI phase noise influence in (7) (full correlation between phase noise samples in time) and (8)-(9) (no correlation between phase noise samples in time). With this normalization we will observe the phase noise influence relative to that of a single channel QPSK system with an RF carrier with a frequency separation of *1/T* where *T* denotes the symbol time. We consider an OFDM system implementation with 4PSK (QPSK) channel modulation.

It is appropriate to evaluate (7)-(9) for all combinations of constellations between the OFDM channels (considering for QPSK channel modulation 4 different constellations per channel) and for all demodulated channels (for all *k*-values). We note that for *N* OFDM channels this leads to an evaluation of $(N-1)\cdot 4^{(N-1)}$ cases for a full investigation and this quickly renders the practical evaluation impossible for increasing *N.* In our specification of a reasonable worst case in the following we have tested a few cases based upon physical intuition (i.e. the same symbol in all channels, a few simple distributions of different symbols in all channels, different received channels in the edges and moving to the center of the OFDM spectrum).

Fig. 3 shows the results as a function of the number of OFDM channels, *N*, for a received OFDM frame where all symbols $a_r$ $(r=0,1,..,N-1,\ r\neq(N-1)/2)$ are the same, and results are shown for the received channel number 0 $(k=0)$. This represents according to our simulation results a worst case for the phase noise influence. Selecting different constellations for the symbols in the OFDM frame and by selecting different received channels, it is possible for all *N*-values to obtain the normalized phase noise variances between 0 and the worst case value shown in the figure. In the case of no correlation between the time samples we find in Fig. 3 the same results using the exact (9) and the more approximate (8). This indicates that the approximation assumption that all phase noise samples are fully correlated between different OFDM channels is a reasonable one. It is thus appropriate to use the approximate (9) for practical system specification. This may speed-up the evaluation since (8) requires the order of $O(N^2)$ computational steps whereas (9) requires $O(N^3)$ steps. In Fig. 3 we see that increasing time-correlation causes increasing phase noise influence. For an *N*-channel CO-OFDM system it is of interest to note that the normalized worst case influence (on the variance) is *N/2* in the case of full time correlation whereas it is *N/4* in the case of no time correlation.

We will investigate the validity of the results in Fig. 3 in some detail. We evaluate the normalized phase noise variance for all constellation configurations and all received channel positions in the OFDM grid for the most important practical design case – the partly correlated case considered in (8)-(9). We do that for *N=5, 7, 9* and *11* and display the results in Fig. 4 in bar diagram format. From Fig. 4 it is clearly observed that system design based on a normalized phase noise variance of *N/4* (as used in Fig. 3) represents a sensible worst case for the selected *N*-values. We tentatively extract this observation to cover all larger *N*-values as well (where the results of Fig. 4 cannot be generated due to the huge amount of $(N-1)\cdot 4^{(N-1)}$ required evaluation cases) and also assume – in accordance with the results of Fig. 4 - that normalized phase noise variance of *N/2* for the fully correlated case (see (7) and Fig. 3) is reasonable as a worst case system design scenario.

We conclude that the results of Fig. 3 represent sensible worst case design guidelines for practical CO-OFDM systems.

We will now move to more detailed practical CO-OFDM system examples. We consider a normal transmission fiber ($D=16$ psec/nm/km) for the distances up to around 500 km, a transmission wavelength of $\lambda = 1.55\ \mu m$, $c = 3\cdot 10^8\ m/sec$, an OFDM channel separation of $\Delta f = 1$ GHz, i.e. baud rate 1 GS/s (symbol time $T=1$ nsec), channel modulation as QPSK, and the number of channels *N* of 101 and 201. (One OFDM channel position in the center of the OFDM grid is used to transmit the RF pilot tone.) For our 100 channel OFDM system case, we have $T=10^{-9}$ sec and (for $L=100$ km) we have $\tau = 0.013\cdot 10^{-9}$ sec. This indicates that the phase noise variance derivation in the case of no time correlation between phase noise samples (i.e. for $T>>\tau$) represents a reasonable choice for the specification of practical CO-OFDM systems that will be used in the access or metro telecom/datacom network with transmission distances below the order of 500-1000 km.

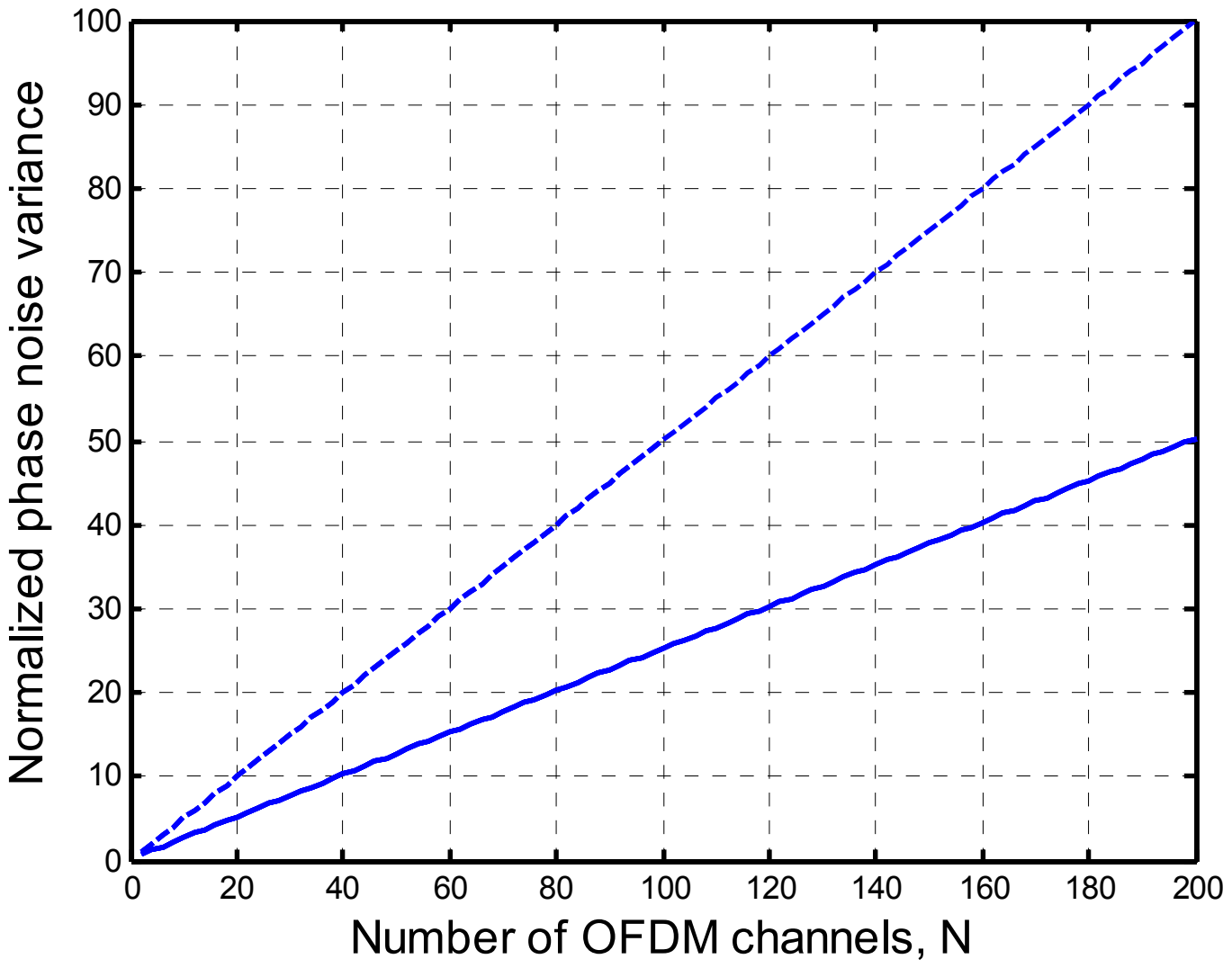

**Figure 3.** Normalized phase noise variance $\sigma^2_{k,CPE+ICI} / 2\pi\Delta\nu_{Tx}\tau$ as a function of the number of OFDM channels $N$ for received channel $k=0,N-1$. Two solid curves (on top of each other) shows results in the case of no time-correlation between phase noise samples in time using (8)-(9); dashed curve shows results in the fully correlated case using (7).

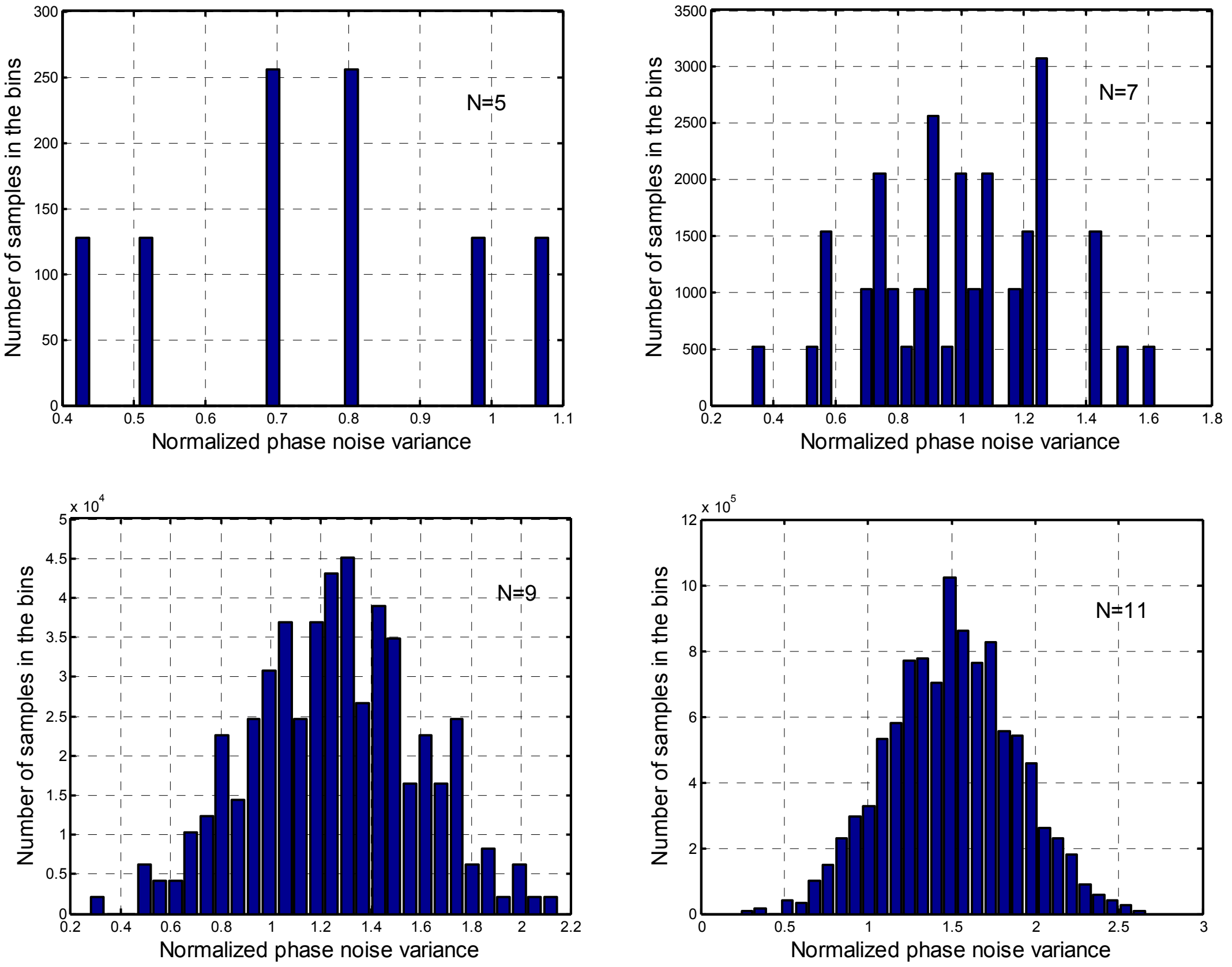

**Figure 4.** Number of samples in a bin representation versus normalized phase noise variance $\sigma^2_{k,CPE+ICI} / 2\pi\Delta\nu_{Tx}\tau$ using (8)-(9) i.e. in the case of no time-correlation between phase noise samples. Number of OFDM channels considered are N=5, 7, 9 and 11 (as indicated) and all constellation configurations and all received channels are considered.

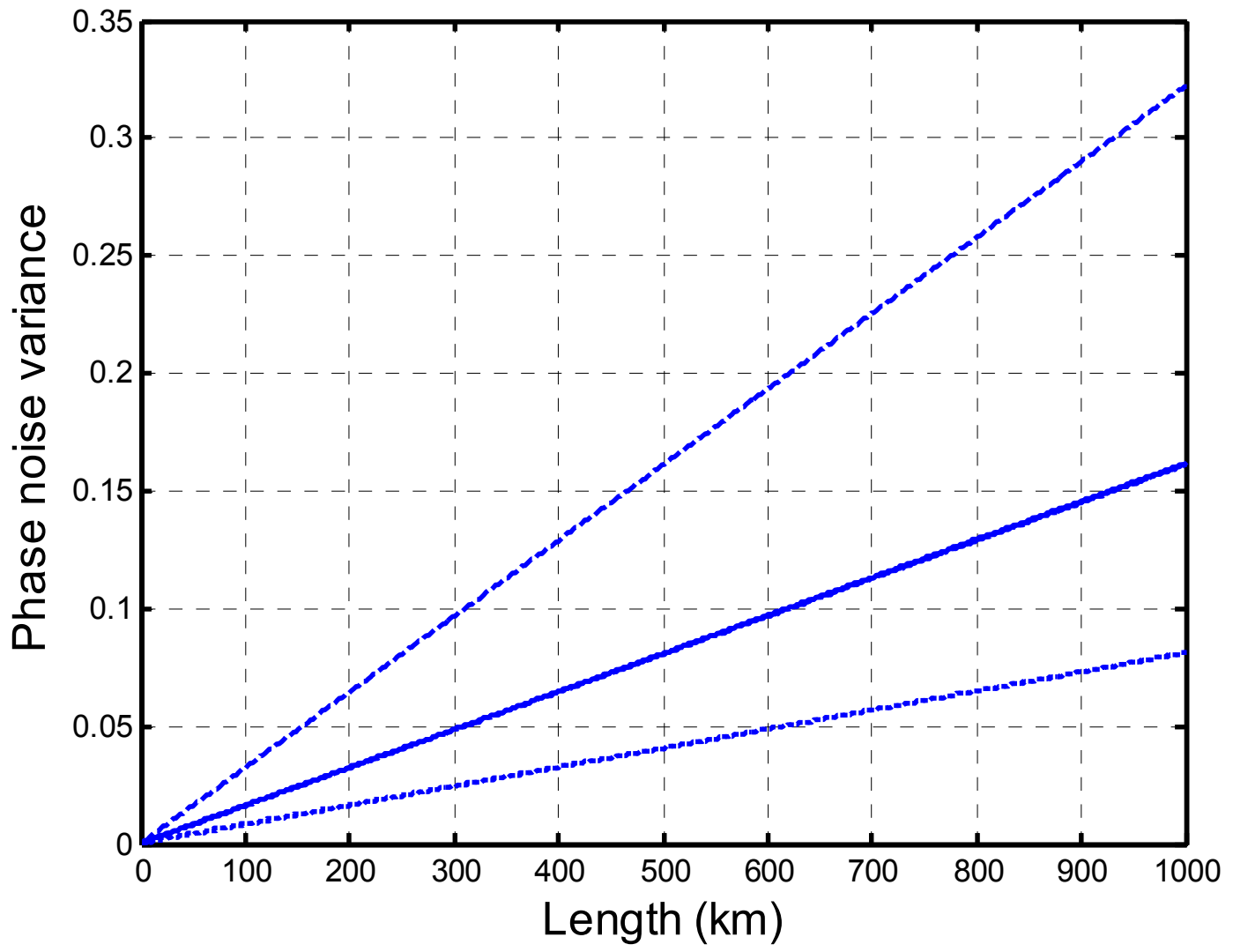

**Figure 5.** Phase noise variance as a function of transmission length. Dashed curve shows results in the case of full correlation between phase noise samples in time using (7) for 200 OFDM channels. Solid curve (in reality three curves on top of each other) show results in three cases 1) for 200 OFDM channels in the case of no time correlation using (8); 2) for 100 OFDM channels for full time correlation, (7); 3) for a 200 GS/s QPSK system using (10). Dotted curve is for a 100 channel OFDM system without time correlation (Eq. (8)) and for a 100 GS/s QPSK system, (10).

We select a Tx linewidth $\Delta\nu_{Tx}$ of 4 MHz, representative for a typical quality DFB laser diode, in the practical evaluation of the CO-OFDM system performance. Figure 4 shows $\sigma^2_{k,CPE+ICI}$ as a function of the transmission length $L$ for the 100 and 200 channel OFDM system. The phase noise variance is compared to the phase noise variance of a single polarization 100 GS/s and 200 GS/s QPSK system which have the same capacity as the OFDM systems. For the 100 and 200 GS/s QPSK system, we consider electronic CD compensation and no RF pilot tone is used for the phase noise compensation. In this case, the phase noise variance is influenced by EEPN, and it is given as [21]:

$$\sigma^2_{QPSK} = 2\pi(\Delta\nu_{Tx} + \Delta\nu_{LO})\cdot T_s + \frac{\pi\lambda^2}{2c}\cdot\frac{D\cdot L\cdot\Delta\nu_{LO}}{T_s} \equiv 2\pi(\Delta\nu_{Tx} + \Delta\nu_{LO} + \Delta\nu_{EEPN})T_s \tag{10}$$

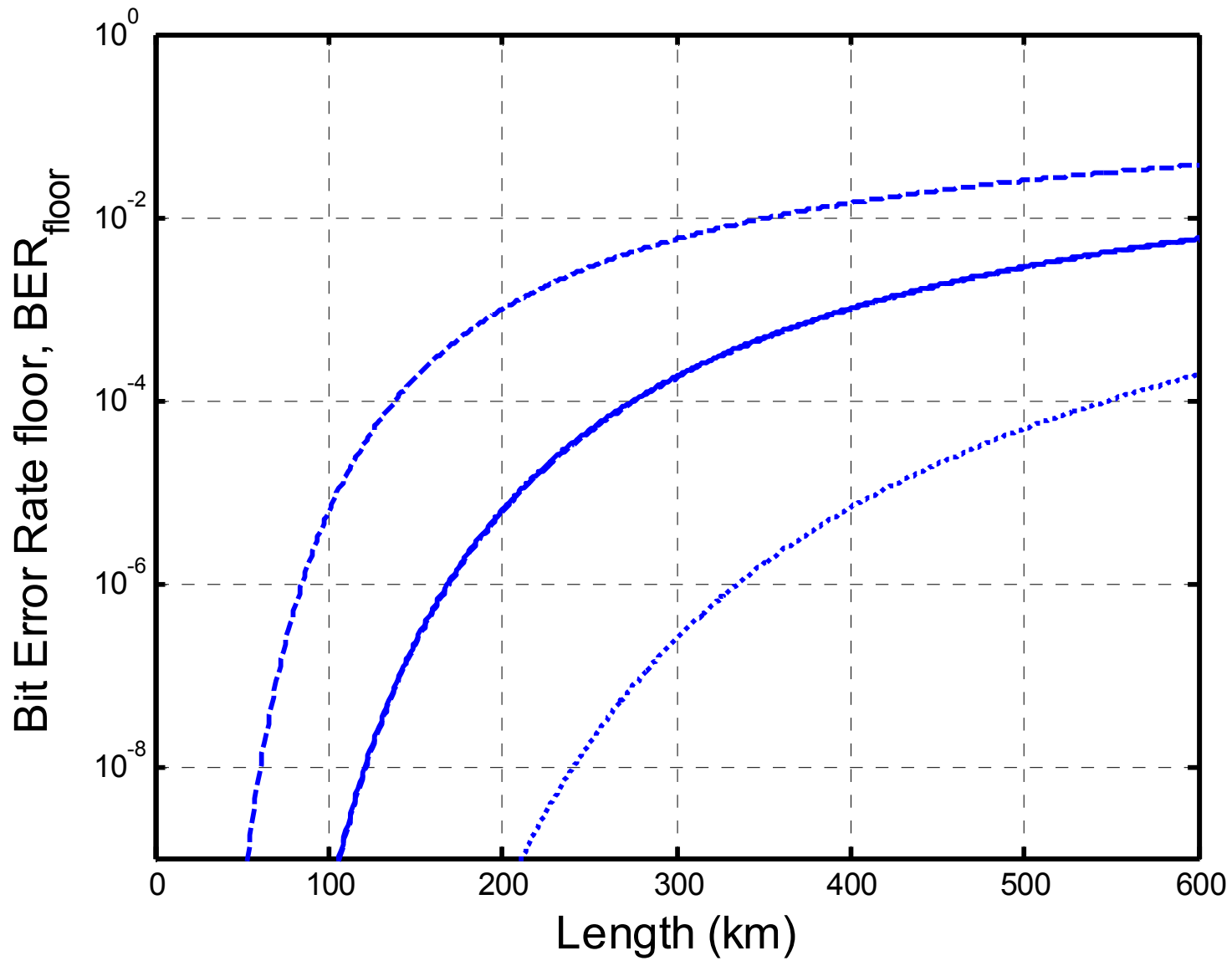

**Figure 6.** Bit-error-rate floor as a function of transmission length. OFDM system performance is shown by dashed curve (200 channels, full time correlation), 3 solid curves on top of each other (200 channels, no correlation in time, 100 channels, full correlation, 200 GS/s QPSK system) and 2 dotted curves on top of each other (100 channels with no correlation in time, 100 GS/s QPSK system).

where $T_s$ is the symbol time which equals $10^{-11}$ sec ($0.5\cdot 10^{-11}$ sec) for the 100 (200) GS/s system. The Transmitter and Local Oscillator linewidths are selected as $\Delta\nu_{Tx} = \Delta\nu_{LO} = 4$ MHz. (13) shows that for the 100 (200) GS/s QPSK system we have for $L = 100$ km the EEPN linewidth $\Delta\nu_{EEPN} = 32\cdot\Delta\nu_{LO}$ $(64\cdot\Delta\nu_{LO})$.

According to results of Fig. 5, the phase noise variance in the case of no time correlation is believed to represents sensible design guidelines for the OFDM system. For completeness we also show the phase noise variance in the case of full time correlation. It is noteworthy that the phase noise variance of the high capacity QPSK system equals that of the worst case OFDM system specification in the case of no time correlation and for systems with the same accumulated capacity. For both type of systems, the phase noise variance increases very much with increasing the transmission distance. For the high capacity QPSK implementation, the increase is due to EEPN whereas for the OFDM implementation, it is due to the dispersion caused walk-off of the RF pilot tone relative to the OFDM signal channels. For an RF pilot located in the center of the OFDM band in a CO-OFDM implementation with nPSK channel modulation, the current results suggest that the walk-off effect is equivalent to the EEPN effect in a single channel nPSK system of the same capacity.

The phase noise parameter of interest (see Fig. 5) are specified by (8) and (10), and may, in general form, be denoted σ. The BER floor for the two 200 Gb/s system implementations is given as [21]:

$$BER_{floor} \approx \frac{1}{2} erfc\left( \frac{\pi}{4\sqrt{2}\sigma} \right) \tag{11}$$

In Figure 6, we display the $BER_{floor}$ versus transmission distance for the phase noise variance cases shown in Fig. 4. A reasonable practical system design constraint is that the BER floor should be below $10^{-4}$ in order for Forward Error Correction (FEC) to operate well. It can be seen that the OFDM systems with capacities of 200, and 400 Gb/s fulfill this requirement for $L < 548$ and 277 km (using the results for no time correlation).

A more approximate analysis of the worst case phase noise sensitivity for analogue CO-OFDM systems with an RF pilot tone and matched filter detection was performed in [15]. In that analysis, it was not identified and pointed out that the RF pilot tone based cancellation of the phase noise will cancel the LO laser phase noise entirely. Keeping this in mind, the results for CO-OFDM systems with the IF linewidth specified is equivalent to the current model with the Tx laser linewidth specified. The 200 channel OFDM system discussed in connection with Figs. 4 and 5 was also considered in [15], and for a 4 MHz linewidth a transmission distance of 225 km was found in order to assure that the BER < $10^{-4}$. This should be compared to the current specification of 277 km link using a more rigorous derivation. Thus, the model in this paper shows good agreement with the earlier more approximate specification. Since the theoretical derivation in section 2 shows that the phase noise sensitivity of the analogue OFDM system with matched filter detection is closely the same as for the software based OFDM system with FFT demodulation, we conclude that the current model framework is the most suitable for estimating the phase noise influence for both systems.

# 4. Conclusions

We present a comparative study of the influence of dispersion induced phase noise for CO-OFDM systems using 1) analogue hardware based channel multiplexing in the Tx and a matched filter Rx; and 2) software based FFT multiplexing and IFFT demultiplexing techniques. For both systems, an RF carrier pilot tone is used to mitigate the phase noise influence. This is, to our knowledge, the first detailed and rigorous study of these two OFDM system configurations. From the analysis it appears that the phase noise influence for the two OFDM implementations is similar. It can be also seen that the theoretical formulation for the software based system provides a method for a rigorous evaluation of the phase noise variance caused by Common Phase Error (CPE= and Inter-Carrier Interference (ICI), and this, in turns, leads to a BER specification.

A major novel theoretal result specifies in exact form - in the limit where the RF pilot tone phase noise cancellation works well - the resulting phase noise variance accounting for the combined CPE and ICI influence including the partial correlation between ICI phase noise samples of different OFDM channels.

The obtained numerical results for the phase noise influence are in agreement with an earlier (more approximate) formulation for analogue hardware based OFDM system [15].

The numerical results of the current study focus on a worst case specification for a CO-OFDM system with 1GS/s QPSK channel modulation. BER results are evaluated and compared to the BER of a QPSK system of the same capacity as the OFDM implementation. Results are evaluated as a function of transmission distance, and for the QPSK system the influence of equalization enhanced phase noise (EEPN) is included. For both type of systems, the phase noise variance increases very much with increasing the transmission distance and the two types of systems have closely the same BER as a function of transmission distance for the same capacity. For the high capacity QPSK implementation the increase in BER is due to EEPN, whereas for the OFDM implementation it is due to the dispersion caused walk-off of the RF pilot tone relative to the OFDM signal channels. For a total capacity of 200 and 400 Gb/s, the transmission distance to have the BER < $10^{-4}$ is less than 548 and 277 km, respectively.

For an RF pilot placed in the center of the OFDM band in a CO-OFDM implementation with nPSK channel modulation, the current results suggest that the walk-off effect is equivalent to the EEPN effect in a single channel nPSK system of the same capacity. This observation is important for future design of coherent long-range systems since it shows that there is a free choice between a CO-OFDM and a high capacity nPSK implementations as far as the phase noise influence is concerned.